\documentclass[12pt]{article}
\usepackage{amsmath}
\usepackage{amssymb}
\usepackage{graphicx}
\usepackage[numbers,sort&compress]{natbib}
\title{Kuznetsov-Ma solution and Akhmediev breather for TD equation
}
\author{Junyi Zhu and Linlin Wang\\
{\small  School of Mathematics and Statistics, Zhengzhou University,}\\
{\small Zhengzhou, Henan 450001, China}\\}
\date{}
\begin{document}
\maketitle
\begin{abstract}
We present the inverse scattering transform for the TD equation, which is related to
the Heisenberg spin equation. We note that the TD equation is an integrable model with high nonlinearity,
and has singularity at zero. Thus a nonzero boundary condition is introduced in the direct scattering problem.
We obtain the $N$-solitonic solution of the TD equation, and give the Kuznetsov-Ma type solution,
rational solution and Akhmediev type breather.
\end{abstract}
\section{Introduction}
The nonlinear integrable equations with nonzero boundary conditions have recently attracted significant interest
in the field of physics and mathematics. The reason is that the localization dynamics of their solitonic solutions are
closely related to the freak (or rogue, extreme) waves in ocean and optic fibres.
Nonlinear Schr\"odinger (NLS) equation (with nonzero boundary condition) is a central model of nonlinear science
\cite{spj37-823,jmp22-2780,non27-r1,F-T1987,smd13-1468,bjp5-337}.
Nowadays several integrable models on the nonzero background have been developed
\cite{ip8-889,jpsj44-1968,ps40-227,jpa36-1931,pre69-066604,jmp47-063508,sam126-245,cmp348-475,jmp51-093506,jmp57-083510,ip23-1711},
and the work of finding another such models is still going on.

The inverse scattering transform method is one of the most important tools in the study of the nonlinear
integrable theory. It is also an efficient way to find the freak waves.
In general, the spectral analysis of the integrable model with nonzero boundary conditions is
much more complicated and difficult than that of the vanishing boundary conditions problem.
One reason may be that it is a result of the modulation instability,
also known as the Benjamin-Feir instability \cite{prsla299-59,jfm27-417}.
While, it is a point of common agreement that the modulation
instability is a possible mechanism for the generation of freak waves.

In the paper, we present an inverse scattering approach to obtain the Kuznetsov-Ma solution \cite{spd22-507,sam60-43}
and Akhmediev breather \cite{spj62-894} for the TD (or Tu-Meng) equation
with high nonlinearity \cite{amas5-89}
\begin{equation}\label{a1}
\begin{aligned}
&iu_t+\left(\frac{v_{xx}}{4v}+u^2-\frac{v^2}{2}\right)_x=0,\\
&iv_t+u_xv+2uv_x=0.
\end{aligned}
\end{equation}
We consider the TD equation with the following non-zero boundary conditions
\begin{equation}\label{b1}
u\to0,\quad v\to\rho_\pm=\pm i\rho,\quad x\to\pm\infty,
\end{equation}
where $u=u(x,t), v=v(x,t)$ and $\rho$ is a positive constant.

Equation (\ref{a1}) is the compatibility condition of the following linear system
\begin{equation}\label{a2}
\psi_x=U\psi,\quad U=i\lambda\sigma_3+Q,\quad Q=u\sigma_3+v\sigma_1,
\end{equation}
and
\begin{equation}\label{a3}
\psi_t=V\psi,\quad V=i\lambda^2\sigma_3+(\lambda
v+iuv)\sigma_1+\frac{v_x}{2}\sigma_2+i\left(u^2+\frac{v_{xx}}{4v}\right)\sigma_3,
\end{equation}
where $\lambda$ is a spectral parameter, $\sigma_j,~ j=1,2,3$ are the classical Pauli matrices.
It is noted that TD equation (\ref{a1}) implies the following conservation laws
\begin{equation}\label{a0}
\begin{aligned}
&iu_t+\left(u^2+\frac{1}{8}(\ln v^2)_{xx}+\frac{1}{16}\left((\ln v^2)_x\right)^2-\frac{v^2}{2}\right)_x=0,\\
&\qquad\qquad \frac{1}{2}i(v^2+\rho^2)_t+(uv^2)_x=0.
\end{aligned}
\end{equation}

We note that there is a Miura transformation between the TD equation and the Heisenberg spin equation \cite{cpl23-1}.
The Hamiltonian structure of the TD hierarchy was discussed by the method of trace identity \cite{amas5-89}.
The algebro-geometric solution to a new (2+1)-dimensional integrable equation associated the TD spectral problem was obtained in \cite{ncb117-925}.
The algebro-geometric solution to the TD hierarchy is given in \cite{mpag16-229}.

The paper is organized as follows. In section 2, we consider the spectral analysis involving the transformation of
spectral parameter, the symmetry condition and the asymptotic behaviors. In section 3,
we construct the associated Riemann-Hilbert problem about the meromorphic function,
and establish the relation between the potential and the scattering data.
by introducing the projection of meromorphic function.
In section 4, we consider the case of reflectionless potential and obtain the explicit solutions of the TD equation
including Kuznetsov-Ma solution and Akhmediev breather.

\setcounter{equation}{0}
\section{Spectral analysis}
In this section, we give the spectral analysis of the spectral problem (\ref{a2}) with boundary conditons (\ref{b1}), that is
\begin{equation}\label{b2}
U\to U_{\pm}=\left(\begin{matrix}
i\lambda& \rho_\pm\\
 \rho_\pm&-i\lambda
\end{matrix}\right), \quad x\to\pm\infty.
\end{equation}
Since the eigenvalues of $U_\infty$ are doubly branched, we introduce an affine parameter $k$ defined by
\begin{equation}\label{b4a}
\lambda=\frac{1}{2}\left(k-\frac{\rho^2}{k}\right),
\end{equation}
and an invertible matrix $M$ as following
\begin{equation}\label{b3}
U_\pm=i\zeta M_\pm\sigma_3M_\pm^{-1},\quad M_\pm=I-\frac{\rho_\pm}{k}\sigma_2,
\end{equation}
where the function
\begin{equation}\label{b4}
\zeta=\frac{1}{2}\left(k+\frac{\rho^2}{k}\right),
\end{equation}
is a single-valued function of $k$.
It is noted that the eigenvalues of the matrix $U_\infty$ are doubly branched (the branch points at $\lambda=\pm i\rho$),
and the Joukowsky transform (\ref{b4a}) map the $\lambda$-plane with a cut as $i[-\rho,\rho]$ into the the $k$-plane.
The branch cut on either sheet is mapped into the circle $C_\rho=\{k\in{\mathbb{C}}:|k|=\rho\}$ \cite{jmp47-063508,jmp55-031506}.
In the following, we discuss the spectral analysis of
the eigenfunctions on the $k$ plane.

In (\ref{a3}),  by the condition (\ref{b1}), we have $V\to V_\infty=\lambda U_\infty, |x|\to\infty$, which means that
$V_\infty$ and $U_\infty$ share the same eigenvectors.
We introduce the matrix solutions $\psi_\pm(x,t,k)$ of the system (\ref{a2}) and (\ref{a3}), such that
\begin{equation}\label{b5}
\begin{aligned}
\psi_\pm(x,t,k)\to E_\pm, \quad x\to\pm\infty,
\end{aligned}
\end{equation}
where $E=E(x,t,k)$ is defined as
\begin{equation}\label{b6}
E_\pm=M_\pm{\rm e}^{i\theta(x,t,k)\sigma_3}, \quad \theta(x,t,k)=\zeta(k)(x+\lambda(k)t).
\end{equation}
There exists the scattering matrix $S(k)$ such that
\begin{equation}\label{b7}
\psi_+(x,t,k)=\psi_-(x,t,k)S(k),\quad S(k)=\left(\begin{matrix}
a(k)&-\tilde{b}(k)\\
b(k)&\tilde{a}(k)
\end{matrix}\right),
\end{equation}
where $S(k)$ is independent of $x,t$.
We note that the different column vectors of the eigenfunctions $\psi_\pm(x,t,k)$ are sectionally holomorphic in the complex $k$ plane.
So, equation (\ref{b7}) can be regarded as a jump condition of a Riemann-Hilbert problem.
For the linear spectral problem (\ref{a2}), the matrix trace of $U$ is zero, which implies that
$\det\psi_\pm(x,k)$ is independent of $x$. Thus, we have
\begin{equation}\label{b8}
\det\psi_\pm(x,t,k)=\det M_\pm=1+\frac{\rho^2}{k^2},\quad \det S(k)=1,
\end{equation}
in terms of (\ref{b5}) and (\ref{b7}).

We now discuss the symmetry condition. According to the definitions (\ref{b4a}) and (\ref{b4}), we find that
$\lambda(-\rho^2/k)=\lambda(k), \zeta(-\rho^2/k)=-\zeta(k)$ and $U(-\rho^2/k)=U(k), V(-\rho^2/k) = V(k)$, and then
\begin{equation}\label{b8a}
\psi_\pm(k)=-\frac{\rho_\pm}{k}\psi_\pm\big(-\frac{\rho^2}{k}\big)\sigma_2.
\end{equation}
Furthermore, for the scattering matrix, we have the following symmetry condition
\begin{equation}\label{b8b}
S\big(-\frac{\rho^2}{k}\big)=-\sigma_2S(k)\sigma_2,
\end{equation}
which implies that the scattering coefficients admit
\begin{equation}\label{b8c}
\tilde{a}(k)=-a(-\frac{\rho^2}{k}\big), \quad \tilde{b}(k)=-(-\frac{\rho^2}{k}\big).
\end{equation}
Here, the relation $\rho_+=-\rho_-=i\rho$ is used.

For the sake of convenience, we introduce the Jost functions $J_\pm(x,k)$ as
\begin{equation}\label{b9}
\psi_\pm(x,k)=J_\pm(x,k){\rm e}^{i\theta\sigma_3}, \quad \theta=\theta(x,k).
\end{equation}
Here and after, we omit the dependence of the time variable $t$.
Then the Jost functions $J_\pm(x,k)$ satisfy
\begin{equation}\label{b10a}
J_x(x,k)+i\zeta(k)J(x,k)\sigma_3=i\lambda(k)\sigma_3J(x,k)+QJ(x,k),
\end{equation}
and
\begin{equation}\label{b10b}
J_t(x,k)+i\zeta(k)\lambda(k)J(x,k)\sigma_3=VJ(x,k).
\end{equation}
Equation (\ref{b10a}) can be rewritten as
\begin{equation}\label{b10}
\begin{aligned}
M_\pm^{-1}J_{\pm,x}(x,k)-\zeta(k) [\sigma_3,M_\pm^{-1}J_\pm(x,k)]=M_\pm^{-1}\Delta Q_\pm J_\pm(x,k),\\
 \Delta Q_\pm=u\sigma_3+(v-\rho_\pm)\sigma_1,
\end{aligned}
\end{equation}
and the asymptotic condition
\begin{equation}\label{b11}
J_\pm(x,k)\to M_\pm=I-\frac{\rho_\pm}{k}\sigma_2,\quad x\to\pm\infty.
\end{equation}
Then we have the following integral equations
\begin{equation}\label{b12}\begin{aligned}
J_\pm(x,k)=M_\pm+\int_{\pm\infty}^x M_\pm{\rm e}^{i(x-\xi)\zeta\hat\sigma_3}M_\pm^{-1}\Delta Q_\pm J_\pm(\xi,k){\rm d}\xi,\\
\end{aligned}
\end{equation}
where ${\rm e}^{\alpha\hat\sigma_3}B={\rm e}^{\alpha\sigma_3}B{\rm e}^{-\alpha\sigma_3}$.
We denote $J_\pm^{[j]}$ as the $j$-th column of the eigenfunction $J_\pm$, and assume that for all $t>0$,
${\bf u}_-\in L^1(-\infty,a)$ and ${\bf u}_+\in L^1(b,\infty)$, for all $a,b\in{\mathbb{R}}$, where ${\bf u}_\pm=(u,v-\rho_\pm)$ \cite{jmp47-063508,sam126-245,jmp55-031506}.
It is shown that $J_+^{[1]}$ and $J_-^{[2]}$ can be analytically extended in the region $D^+$,
$J_-^{[1]}$ and $J_+^{[2]}$ can be analytically extended in the region $D^-$, where the regions $D^\pm$ are
defined as following
\begin{equation}\label{b13}
\begin{aligned}
D^+=\{k\in{\mathbb{C}}: (|k|^2-\rho^2){\rm Im}k>0\},\quad
D^-=\{k\in{\mathbb{C}}: (|k|^2-\rho^2){\rm Im}k<0\}.
\end{aligned}
\end{equation}
Their boundary curve is denoted by $\partial D=\{k\in{\mathbb{C}}: (|k|^2-\rho^2){\rm Im}k=0\}$.
In fact, for example, if we take $w(x,k)=(M^{-1}J_-)^{[1]}(x,k)$, then equation (\ref{b12}) reduces to
\begin{equation}\label{b14}
w(x,k)=\left(\begin{array}{c}
1\\
0
\end{array}\right)+\int_{-\infty}^xC_-(x,y,k)w(y,k){\rm d}y,
\end{equation}
where $C_-(x,y,k)={\rm diag}(1,{\rm e}^{-2i(x-y)\zeta})M_-^{-1}\Delta Q_-M_-$. By virtue of the $L^1$ vector norm $\|w\|=|w_1|+|w_2|$
and the corresponding subordinate norm $\|M_-\|=1+\rho/|k|$ and $\|M_-^{-1}\|=(1+\rho/|k|)/|1+\rho^2/k^2|$, one finds
$\|C_-(x,y,k)\|\leq 2 \rho_\epsilon (|u|+|v-\rho_-|), k\in D^-_\epsilon:=\overline{D^-}\setminus B_\epsilon(\pm i\rho).$
Here $\rho_\epsilon={\rm Max}\{\|M_-\|\|M_-^{-1}\|\}, k\in D^-_\epsilon$, with $B_\epsilon(\pm i\rho)=\{k\in{\mathbb{C}}:|k\mp i\rho|<\epsilon\rho\}$.
Hence, it is shown, by induction, that $w(x,k)$ can be expanded as the following neumann series as
\[w(x,k)=\sum\limits_{n=1}^\infty w^{(n)}(x,k), \quad \|w^{(n)}(x,k)\|\leq\frac{\mu^n(x)}{n!},\]
where $\mu(x)=2\rho_\epsilon\int_{-\infty}^x(|u|+|v-\rho_-|){\rm d}y$ and
\[w^{(0)}=\left(\begin{array}{c}
1\\
0
\end{array}\right),\quad w^{(n+1)}(x,k)=\int_{-\infty}^xC_-(x,y,k)w^{(n)}(y,k){\rm d}y.\]
Thus, for all $\epsilon>0$, if ${\bf u}_-\in L^1(-\infty,a)$, for all $a\in{\mathbb{R}}$,
the Neumann series converges absolutely and uniformly with respect to $x\in(-\infty,a)$ and $k\in D^-$,
which means that $J_-^{[1]}$ is analytic in the region $D^-$.

Since $\zeta(\pm i\rho)=0$, the exponentials ${\rm e}^{\pm i(x-\xi)\zeta}$ in (\ref{b12}) reduce to the identity,
and the matrix $M_\pm$ are degenerate. So $M_\pm^{-1}(\pm i\rho)$ do not exist. However, we note that
the term $M_\pm(k){\rm e}^{i(x-\xi)\zeta}M_\pm^{-1}(k)$ appearing in (\ref{b12}) remains finite as $k\to\pm i\rho$.
Thus both columns of $J_\pm(x,k)$ remain well-defined at the points $k=\pm i\rho$.
\begin{figure}[h]
\setlength{\unitlength}{0.1in}
\begin{picture}(10,10)
\put(25,5){\line(1,0){10}}
\put(30,5){\circle{40}} \put(29.5,7.8){\vector(1,0){1}}
\put(29.5,2.3){\vector(1,0){1}} \put(31,5){\vector(-1,0){1}}
\put(25.5,5){\vector(1,0){1}} \put(33.5,5){\vector(1,0){1}}
\put(25.3,7){\makebox(2,1)[l]{\scriptsize$D^+$}}
\put(32.5,7){\makebox(2,1)[l]{\scriptsize$D^+$}}
\put(30,3){\makebox(2,1)[l]{\scriptsize$D^+$}}
\put(29.3,6){\makebox(2,1)[l]{\scriptsize$D^-$}}
\put(25.5,2.5){\makebox(2,1)[l]{\scriptsize$D^-$}}
\put(32.5,2.5){\makebox(2,1)[l]{\scriptsize$D^-$}}
\put(29.5,3.9){\makebox(2,1)[l]{\scriptsize$o$}}
\put(26.95,4.4){\makebox(2,1)[l]{\tiny$\bullet$}}
\put(32.45,4.4){\makebox(2,1)[l]{\tiny$\bullet$}}
\put(35.45,4.4){\makebox(2,1)[l]{\scriptsize$\partial D$}}
\put(30.1,7.9){\makebox(2,1)[l]{\scriptsize$i\rho$}}
\put(27.9,1.3){\makebox(2,1)[l]{\scriptsize$-i\rho$}}
\end{picture}
\caption{Domain $D^+$, $D^-$ and their boundary curve $\partial D$}
\end{figure}
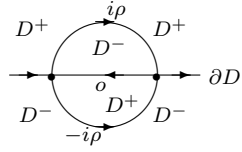

In addition, from (\ref{b7}) and (\ref{b8}), as well as (\ref{b9}), one find that
\begin{equation}\label{b13a}
\begin{aligned}
a(k)&=\frac{1}{ \mathcal{M}}W(J_+^{[1]},J_-^{[2]}), \quad &\tilde{a}(k)=\frac{1}{\mathcal{M}}W(J_-^{[1]},J_+^{[2]}),\\
b(k)&=\frac{1}{\mathcal{M}}{\rm e}^{2i\theta}W(J_-^{[1]},J_+^{[1]}),\quad &\tilde{b}(k)=\frac{1}{\mathcal{M}}{\rm e}^{-2i\theta}W(J_-^{[2]},J_+^{[2]}),
\end{aligned}
\end{equation}
where ${\cal{M}}=\det M_\pm=1+\frac{\rho^2}{k^2}$, and $W(\cdot,\cdot)$ denotes the Wronski determinant. So $a(k)$ is analytic in $D^+$, and $\tilde{a}(k)$ is analytic in $D^-$.
Suppose that $a(k)$ has $N$ simple zeros $\{k_j\}$, ${\rm Im}k_j>0,
j=1,\cdots,n_1; {\rm Im}k_j<0, j=n_1+1,\cdots,n_1+n_2=N$.
Then, the points $\tilde{k}_j=-\rho^2/k_j, j=1,\cdots,N$ are the zeroes of $\tilde{a}(k)$ by the symmetry (\ref{b8c}).

Since $a(k_j)=0$ and $\tilde{a}(\tilde{k}_j)=0$, equation (\ref{b13a}) implies that
\begin{equation}\label{b15}
J_+^{[1]}(k_j)=b_jJ_-^{[2]}(k_j), \quad J_+^{[2]}(\tilde{k}_j)=\tilde{b}_jJ_-^{[1]}(\tilde{k}_j),
\end{equation}
where the proportional coefficients $b_j$ and $\tilde{b}_j$ are the functions of variables $x$ and $t$, that is
\begin{equation}\label{b16}
b_j=b_{j0}{\rm e}^{-2\theta(k_j)}, \quad \tilde{b}_j=\tilde{b}_{j0}{\rm e}^{2i\theta(\tilde{k}_j)}.
\end{equation}
Here, $b_{j0}$ and $\tilde{b}_{j0}$ are constants, and $\theta(k)$ is defined in (\ref{b6}).
From the symmetry condition (\ref{b8a}) and the transformation (\ref{b9}), one finds that $J_\pm(k)$ also satisfy the symmetry
condition (\ref{b8a}), which implies that
$$J_\pm^{[1]}(x,k)=-i\frac{\rho_\pm}{k}J_\pm^{[2]}(x,-\frac{\rho^2}{k}),\quad J_\pm^{[2]}(x,k)=i\frac{\rho_\pm}{k}J_\pm^{[1]}(x,-\frac{\rho^2}{k}).$$
Thus, we have
\begin{equation}\label{b17}
b_j=\tilde{b}_j.
\end{equation}

It is remarked that $a(k)$ and $\tilde{a}(k)$ also have simple poles at $k=\pm i\rho$ on the boundary curve $\partial D$ in view of
${\mathcal{M}}=1+\rho^2/k^2$. Since $\det J_\pm(x,\pm i\rho)=0$, by virtue of the condition (\ref{b8a}), we have
\[J_\pm^{[2]}(x, i\rho)=i J_\pm^{[1]}(x,i\rho),\quad J_\pm^{[2]}(x, -i\rho)=-i J_\pm^{[1]}(x,- i\rho).\]
Furthermore, we have
\begin{equation}\label{b13b}
\begin{aligned}
a(k)=\frac{a_\pm}{k\mp i\rho}+O(1),\quad \tilde{a}(k)=\frac{\tilde{a}_\pm}{k\mp i\rho}+O(1),\\
b(k)=\frac{b_\pm}{k\mp i\rho}+O(1), \quad \tilde{b}(k)=\frac{\tilde{b}_\pm}{k\mp i\rho}+O(1),
\end{aligned}
\end{equation}
where
\begin{equation}\label{b13c}
\begin{aligned}
a_\pm=\pm\frac{i\rho}{2}W(J_+^{[1]}(\pm i\rho),J_-^{[2]}(\pm i\rho))=\mp ib_\pm, \\
\tilde{a}_\pm=\pm\frac{i\rho}{2}W(J_-^{[1]}(\pm i\rho),J_+^{[2]}(\pm i\rho))=\mp i\tilde{b}_\pm.
\end{aligned}
\end{equation}
Moreover, the symmetry condition $\tilde{a}(k)=a(-\rho^2/k)$ implies that $\lim\limits_{k=\pm i\rho}a(-\rho^2/k)/\tilde{a}(k)=1$.
So $\tilde{a}_\pm=a_\pm$.
Note that $k=0$ is not the zero of $a(k)$ and $\tilde{a}(k)$, which will be shown in the end of the section.

we next discuss the asymptotic behaviors of the Jost functions. It is noted that the limit $\lambda\to\infty$ corresponds to
$k\to\infty$ in one sheet and to $k\to0$ in another sheet.
As $k\to\infty$, we let
\begin{equation}\label{b19}
J_\pm(x,k)=J_\pm^{(0)}(x)+\frac{1}{k}J_\pm^{(1)}(x)+\frac{1}{k^2}J_\pm^{(2)}(x)+\cdots,\quad
k\to\infty,
\end{equation}
and
\begin{equation}\label{b29}
J_\pm(x,k)=k^{-1}\hat{J}_\pm^{(-1)}(x)+\hat{J}_\pm^{(0)}(x)+k\hat{J}_\pm^{(1)}(x)+\cdots,\quad
k\to0.
\end{equation}
By substituting them into (\ref{b10a}), and using the boundary condition (\ref{b11}), we obtain
\begin{equation}\label{b23}
J_\pm^{(0)}(x)={\rm e}^{\eta_\pm\sigma_3},\quad
J_\pm^{(1)}(x)=-(v\sigma_2+i\gamma_\pm\sigma_3){\rm e}^{\eta_\pm\sigma_3},
\end{equation}
and
\begin{equation}\label{b30}
\hat{J}_\pm^{(-1)}(x)=\left(\begin{array}{cc}
0&i\rho_\pm{\rm e}^{\eta_\pm}\\
-i\rho_\pm{\rm e}^{-\eta_\pm}&0
\end{array}\right),\quad \hat{J}_\pm^{(0)}(x)=\left(\begin{array}{cc}
\frac{v}{\rho_\pm}{\rm e}^{-\eta_\pm}&\frac{\gamma_\pm}{\rho_\pm}{\rm e}^{\eta_\pm}\\
\frac{\gamma_\pm}{\rho_\pm}{\rm e}^{-\eta_\pm}&\frac{v}{\rho_\pm}{\rm e}^{\eta_\pm}
\end{array}\right),
\end{equation}
where
\[\gamma_\pm=\int_{\pm\infty}^x(\rho^2+v^2(\xi)){\rm d}\xi,\quad \eta^\pm=\int_{\pm\infty}^xu(\xi){\rm d}\xi.\]

Hence, from (\ref{b13a}), we find
\begin{equation}\label{b26}
\begin{aligned}
&a(k)\to{\rm e}^{-\eta}\left(1+\frac{i\gamma}{k}\right),\quad  k\to\infty,\\
&a(k)\to-{\rm e}^{\eta}\big(1+k\frac{i\gamma}{\rho^2}\big), \quad k\to0,
\end{aligned}
\end{equation}
and
\begin{equation}\label{b26b}
\begin{aligned}
&\tilde{a}(k)\to{\rm e}^{\eta}\left(1-\frac{i\gamma}{k}\right),\quad  k\to\infty,\\
&\tilde{a}(k)\to-{\rm e}^{-\eta}\big(1-k\frac{i\gamma}{\rho^2}\big), \quad k\to0,
\end{aligned}
\end{equation}
where
\begin{equation}\label{b25a}
\eta=\eta^--\eta^+=\int_{-\infty}^\infty u(x){\rm d}x, \quad \gamma=\gamma^--\gamma^+=\int_{-\infty}^\infty(\rho^2+v^2(x)){\rm d}x.
\end{equation}
From (\ref{a0}), one finds that $\eta$ and $\gamma$ are constants.

It is remarked that, since $\Delta Q_\pm$ is not an off-diagonal matrix, the asymptotic behaviors of the Jost functions
can not be derived from the integral equation (\ref{b12}) by constructing the iterative series.

\setcounter{equation}{0}
\section{Riemann-Hilbert problem}
From (\ref{b7}) and (\ref{b9}), one finds that
\begin{equation}\label{c1}
J_+(x,k)=J_-(x,k){\rm e}^{i\theta\sigma_3}S(k){\rm e}^{-i\theta\sigma_3}, \quad k\in \partial D.
\end{equation}
Introducing the sectionally meromorphic matrices
\begin{equation}\label{c2}
\Phi^+(x,k)=\left(\frac{J_+^{[1]}(x,k)}{a(k)},J_-^{[2]}(x,k)\right),\quad \Phi^-(x,k)=\left(J_-^{[1]}(x,k),\frac{J_+^{[2]}(x,k)}{\tilde{a}(k)}\right),
\end{equation}
equation (\ref{c1}) gives rises to the following jump condition
\begin{equation}\label{c3}
\Phi^-(x,k)-\Phi^+(x,k)=\Phi^+(x,k)G(x,k),\quad k\in \partial D,
\end{equation}
where the matrix $G(x,k)$ is
\[G(x,k)=\left(\begin{matrix}
0&-\tilde\varrho(k){\rm e}^{2i\theta}\\
-\varrho(k){\rm e}^{-2i\theta}&\varrho(k)\tilde\varrho(k)
\end{matrix}\right), \quad \varrho(k)=\frac{b(k)}{a(k)},\quad \tilde\varrho(k)=\frac{\tilde{b}(k)}{\tilde{a}(k)}.\]

By virtue of the asymptotic properties (\ref{b19})-(\ref{b26b}), it is readily verified that the sectionally meromorphic functions
$\Phi^\pm(x,k)$ admit the following normalization condition
\begin{equation}\label{c4}
\Phi^\pm(x,k)={\rm e}^{\eta_-\sigma_3}+O\big(\frac{1}{k}\big),\quad k\to\infty,
\end{equation}
as well as
\begin{equation}\label{c5}
\Phi^\pm(x,k)=\frac{1}{k}\Phi_{-1}+O\big(1\big),\quad k\to0.
\end{equation}
Here
\begin{equation}\label{c6}
\Phi_{-1}=\left(\begin{matrix}
0&\rho{\rm e}^{\eta_-}\\
-\rho{\rm e}^{-\eta_-}&0
\end{matrix}\right).
\end{equation}

To establish the relation between the potential and the scattering data from
the Riemann-Hilbert problem (\ref{c3})-(\ref{c5}), we introduce the Cauchy projectors ${\mathcal{P}}_\pm$
over the oriented contour $\partial D$ shown in figure 1,
\[{\mathcal{P}}_\pm[f](k)=\frac{1}{2\pi i}\int_{\partial D}\frac{f(z)}{z-(k\pm i0)}{\rm d}z,\]
where the notation $k\pm i0$ indicates that when $k\in\partial D$, the limit is taken from the left/right of it.
For the sectionally meromorphic functions $f^\pm(k)$, with poles $z_j^\pm\in D^{\pm}$, if $f^\pm(k)\to0$ as $k\to\infty$,
then
\begin{equation}\label{c7}
\begin{aligned}
{\mathcal{P}}_\pm[f^\pm](k)=&\mp\sum\limits_j\frac{{\rm Res}[f^\pm,z_j^\pm]}{k-z_j^\pm}\pm f^\pm(k),\\
{\mathcal{P}}_\mp[f^\pm](k)=&\mp\sum\limits_j\frac{{\rm Res}[f^\pm,z_j^\pm]}{k-z_j^\pm}.
\end{aligned}
\end{equation}
Note that $\Phi^\pm(x,k)-{\rm e}^{\eta_-\sigma_3}-\Phi_{-1}/k$ are sectionally meromorphic functions, with poles $k_j\in D^+$ and
$\tilde{k}_j\in D^-, (j=1,\cdots,n)$, and tend to zero as $k\to\infty$. Thus
\[{\mathcal{P}}_-[\Phi^-(x,k)-{\rm e}^{\eta_-\sigma_3}-\frac{\Phi_{-1}}{k}]-{\mathcal{P}}_-[\Phi^+(x,k)-{\rm e}^{\eta_-\sigma_3}-\frac{\Phi_{-1}}{k}]={\mathcal{P}}_-[\Phi^+(x,k)G(x,k)],\]
implies that
\[\begin{aligned}
\left(\sum\limits_{j=1}^N\frac{{\rm Res}[\Phi^-,\tilde{k}_j]}{k-\tilde{k}_j}-[\Phi^-(x,k)-{\rm e}^{\eta_-\sigma_3}-\frac{\Phi_{-1}}{k}]\right)+\sum\limits_{j=1}^N\frac{{\rm Res}[\Phi^+,k_j]}{k-k_j}\\
={\mathcal{P}}_-[\Phi^+(x,k)G(x,k)].
\end{aligned}\]
Similarly, by applying ${\mathcal{P}_+}$, we have
\[\begin{aligned}
\sum\limits_{j=1}^N\frac{{\rm Res}[\Phi^-,\tilde{k}_j]}{k-\tilde{k}_j}-\left(-\sum\limits_{j=1}^N\frac{{\rm Res}[\Phi^+,k_j]}{k-k_j}+[\Phi^+(x,k)-{\rm e}^{\eta_-\sigma_3}-\frac{\Phi_{-1}}{k}]\right)\\
={\mathcal{P}}_+[\Phi^+(x,k)G(x,k)].
\end{aligned}\]
Thus, we have
\begin{equation}\label{c8}
\begin{aligned}
\Phi^\pm(x,k)={\rm e}^{\eta_-\sigma_3}+\frac{\Phi_{-1}}{k}+\sum\limits_{j=1}^N\frac{{\rm Res}[\Phi^+,k_j]}{k-k_j}+\sum\limits_{j=1}^N\frac{{\rm Res}[\Phi^-,\tilde{k}_j]}{k-\tilde{k}_j}\\
-{\mathcal{P}}_\pm[\Phi^+(x,k)G(x,k)],
\end{aligned}
\end{equation}
which implies that
\begin{equation}\label{c9}
\begin{aligned}
\Phi^\pm(x,k)={\rm e}^{\eta_-\sigma_3}+\frac{\Phi_{-1}}{k}+\frac{1}{k}\sum\limits_{j=1}^N\left({\rm Res}[\Phi^+,k_j]+{\rm Res}[\Phi^-,\tilde{k}_j]\right)\\
+\frac{1}{k}\frac{1}{2\pi i}\int_{\partial D}[\Phi^+(x,k)G(x,k)]{\rm d}k, \quad k\to\infty.
\end{aligned}
\end{equation}

We note that the residues ${\rm Res}[\Phi^+,k_j]$ and ${\rm Res}[\Phi^-,\tilde{k}_j]$ take the following form
\begin{equation}\label{c11}
\begin{aligned}
{\rm Res}[\Phi^+,k_j]=(g_jJ_-^{[2]}(x,k_j),0),\\
{\rm Res}[\Phi^-,\tilde{k}_j]=(0,\tilde{g}_jJ_-^{[1]}(x,\tilde{k}_j)),
\end{aligned}
\end{equation}
where
\begin{equation}\label{c12}
g_j=\frac{b_{j0}}{\dot{a}(k_j)}{\rm e}^{-2i\theta(k_j)}, \quad \tilde{g}_j=\frac{\tilde{b}_{j0}}{\dot{\tilde{a}}(\tilde{k}_j)}{\rm e}^{2i\theta(\tilde{k}_j)}.
\end{equation}
Here the dot denotes the derivative with respect to $k$, and $\theta(k)$ is defined in (\ref{b6}).
It is readily verified that
\begin{equation}\label{c13}
\tilde{g}_j=-\frac{\rho^2}{k_j^2}g_j
\end{equation}
by virtue of the symmetry condition (\ref{b8c}) and
the identities $\lambda(-\rho^2/k)=\lambda(k)$, $\zeta(-\rho^2/k)=-\zeta(k)$.

On the other hand, from (\ref{b29})-(\ref{b26b}), one finds
\begin{equation}\label{c14}
\Phi^\pm(x,k)={\rm e}^{\eta_-\sigma_3}+\frac{1}{k}\left(\begin{matrix}
-i\gamma_-&iv\\
-iv&i\gamma_-
\end{matrix}\right){\rm e}^{\eta_-\sigma_3}+O\big(\frac{1}{k^2}\big),\quad k\to\infty.
\end{equation}

Thus, from the $O(1/k)$ term of equations (\ref{c9}) and (\ref{c14}), we find the relation
between the potential and the scattering data as following
\begin{equation}\label{c15}
\begin{aligned}
\left(\begin{matrix}
-i\gamma_-{\rm e}^{\eta_-}&iv{\rm e}^{-\eta_-}\\
-iv{\rm e}^{\eta_-}&i\gamma_-{\rm e}^{-\eta_-}
\end{matrix}\right)=&\left(\begin{matrix}
0&\rho{\rm e}^{\eta_-}\\
-\rho{\rm e}^{-\eta_-}&0
\end{matrix}\right)+\sum\limits_{j=1}^N\left((g_jJ_-^{[2]}(x,k_j),\mathbf{0})+(\mathbf{0},\tilde{g}_jJ_-^{[1]}(x,\tilde{k}_j))\right)\\
&\quad+\frac{1}{2\pi i}\int_{\partial D}[\Phi^+(x,k)G(x,k)]{\rm d}k.
\end{aligned}
\end{equation}

\setcounter{equation}{0}
\section{Reflectionless potentials and solutions}
Now, we consider the case of reflectionless potentials, that is $G(x,k)\equiv0$.
In this case, there is no jump from $\Phi^+$ to $\Phi^-$ across the continuous spectrum, and
the inverse problem reduces to an algebraic system. Therefore, one can obtain the soliton solutions of TD equation.

Equation (\ref{c15}) reduces to
\begin{equation}\label{d1}
\left(\begin{array}{c}
-i\gamma_-\\
-iv
\end{array}\right){\rm e}^{\eta_-}+\left(\begin{array}{c}
0\\
\rho
\end{array}\right){\rm e}^{-\eta_-}=\sum\limits_{j=1}^Ng_jJ_-^{[2]}(x,k_j),
\end{equation}
and
\begin{equation}\label{d2}
\left(\begin{array}{c}
iv\\
i\gamma_-
\end{array}\right){\rm e}^{-\eta_-}-\left(\begin{array}{c}
\rho\\
0
\end{array}\right){\rm e}^{\eta_-}=\sum\limits_{j=1}^N\tilde{g}_jJ_-^{[1]}(x,\tilde{k}_j).
\end{equation}
Here, the functions $J_-^{[1]}(x,\tilde{k}_l)$ and $J_-^{[2]}(x,k_j)$ satisfy the following algebraic system
\begin{equation}\label{d3}
J_-^{[1]}(x,\tilde{k}_l)=\left(\begin{array}{c}
{\rm e}^{\eta_-}\\
0
\end{array}\right)+\frac{1}{\tilde{k}_l}\left(\begin{array}{c}
0\\
-\rho{\rm e}^{-\eta_-}
\end{array}\right)-\sum\limits_{j=1}^NJ_-^{[2]}(x,k_j)\frac{g_j}{k_j-\tilde{k}_l},
\end{equation}
and
\begin{equation}\label{d4}
J_-^{[2]}(x,k_j)=\left(\begin{array}{c}
0\\
{\rm e}^{-\eta_-}
\end{array}\right)+\frac{1}{k_j}\left(\begin{array}{c}
\rho{\rm e}^{\eta_-}\\
0
\end{array}\right)-\sum\limits_{l=1}^NJ_-^{[1]}(x,\tilde{k}_l)\frac{\tilde{g}_l}{\tilde{k}_l-k_j}.
\end{equation}

For the sake of convenience, we introduce the following matrices
\begin{equation}\label{d5}
\begin{aligned}
{\bf J}_-^{[1]}=\left(J_-^{[1]}(x,\tilde{k}_1),J_-^{[1]}(x,\tilde{k}_2),\cdots,J_-^{[1]}(x,\tilde{k}_N)\right),\\
{\bf J}_-^{[2]}=\left(J_-^{[2]}(x,k_1),J_-^{[2]}(x,k_2),\cdots,J_-^{[2]}(x,k_N)\right),\\
{\bf g}=(g_1,g_2,\cdots,g_N),\quad {\bf\tilde{g}}=(\tilde{g}_1,\tilde{g}_2,\cdots,\tilde{g}_N),
\end{aligned}
\end{equation}
and $\mathbf{A}=I-\mathbf{K}\tilde{\mathbf{K}}, \tilde{\mathbf{A}}=I-\tilde{\mathbf{K}}\mathbf{K}$,
\begin{equation}\label{d6}
\begin{aligned}
 \mathbf{K}=(K_{jl}),\quad \tilde{\mathbf{K}}=(\tilde{K}_{lj}),\quad
K_{jl}=\frac{g_j}{k_j-\tilde{k}_l},\quad \tilde{K}_{lj}=\frac{\tilde{g}_l}{\tilde{k}_l-k_j}.
\end{aligned}
\end{equation}
Since $\mathbf{K}, \tilde{\mathbf{K}}$ are Cauchy type matrices \cite{jpa42-404005,jpa46-035204}, and
$\tilde{\mathbf{A}}=\tilde{\mathbf{K}}^{-1}\mathbf{A}\tilde{\mathbf{K}}$,
we get $\det\tilde{\mathbf{A}}=\det\mathbf{A}$.
In addition, we set
\begin{equation}\label{d7}
\begin{aligned}
{\bf f}=\left(\begin{array}{c}
{\bf f}_1\\
{\bf f}_2
\end{array}\right), \quad {\bf f}_m=(f_{m,1},f_{m,2},\cdots,f_{m,N}),\\
{\bf h}=\left(\begin{array}{c}
{\bf h}_1\\
{\bf h}_2
\end{array}\right), \quad {\bf h}_m=(h_{m,1}, h_{m,2},\cdots,h_{m,N}),\\
\end{aligned}
\end{equation}
where $m=1,2$ and these components are given by
\begin{equation}\label{d8}
\begin{aligned}
f_{1,j}=\frac{\rho}{k_j}-\sum\limits_{l=1}^N\tilde{K}_{lj}, \quad f_{2,j}=1+\sum\limits_{l=1}^N\frac{\rho}{\tilde{k}_l}\tilde{K}_{lj},\\
h_{1,l}=1-\sum\limits_{j=1}^N\frac{\rho}{k_j}K_{jl}, \quad h_{2,l}=-\frac{\rho}{\tilde{k}_l}-\sum\limits_{j=1}^NK_{jl}.
\end{aligned}
\end{equation}

Thus, the solution of the algebraic system (\ref{d3}) and (\ref{d4}) is given by
\begin{equation}\label{d9}
{\bf J}_-^{[1]}={\rm e}^{\eta_-\sigma_3}{\bf h}\tilde{\bf A}^{-1}, \quad
{\bf J}_-^{[2]}={\rm e}^{\eta_-\sigma_3}{\bf f}{\bf A}^{-1}.
\end{equation}
Substitution of the above results into the reconstruction formula (\ref{d1}) and (\ref{d2}), we have
\begin{equation}\label{d10}
\gamma_-=-i\frac{\det{\bf A}^{(a)}}{\det{\bf A}}, \quad \gamma_-=i\frac{\det\tilde{\bf A}^{(a)}}{\det{\bf A}},
\end{equation}
and
\begin{equation}\label{d11}
iv{\rm e}^{2\eta_-}=\frac{\det{\bf A}^{(b)}}{\det{\bf A}}, \quad
iv{\rm e}^{-2\eta_-}=\frac{\det\tilde{\bf A}^{(b)}}{\det{\bf A}},
\end{equation}
where the matrices ${\bf A}^{(a)}, {\bf A}^{(b)}$ and $\tilde{\bf A}^{(a)}, \tilde{\bf A}^{(b)}$ are block matrices,
and have the following definitions
\begin{equation}\label{d12}
{\bf A}^{(a)}=\left(\begin{matrix}
0&{\bf f}_1\\
{\bf g}^T&{\bf A}
\end{matrix}\right), \quad {\bf A}^{(b)}=\left(\begin{matrix}
\rho&{\bf f}_2\\
{\bf g}^T&{\bf A}
\end{matrix}\right),
\end{equation}
and
\begin{equation}\label{d13}
\tilde{\bf A}^{(a)}=\left(\begin{matrix}
0&{\bf h}_2\\
{\bf \tilde{g}}^T&\tilde{\bf A}
\end{matrix}\right), \quad \tilde{\bf A}^{(b)}=\left(\begin{matrix}
\rho&-{\bf h}_1\\
{\bf \tilde{g}}^T&\tilde{\bf A}
\end{matrix}\right).
\end{equation}

Finally, the $N$-soliton solution of the TD equation is of the form
\begin{equation}\label{d14}
v^2+\rho^2=-i\partial_x\left(\frac{\det{\bf A}^{(a)}}{\det{\bf A}}\right),\quad
u=\frac{1}{4}\partial_x\ln\left(\frac{\det{\bf A}^{(b)}}{\det\tilde{\bf A}^{(b)}}\right).
\end{equation}

It is remarked that equation (\ref{d10}) implies that $\det{\bf A}^{(a)}+\det\tilde{\bf A}^{(a)}=0$, or equivalently
\begin{equation}\label{d15}
{\bf f}_1{\bf g}^T+{\bf h}_2{\bf \tilde{g}}^T=0.
\end{equation}
We note that equation (\ref{d15}) can be proved by using $\tilde{k}_j=-\rho^2/k_j$ and (\ref{c13}).
In fact, from (\ref{d8}), we find
\begin{equation}\label{d16}
\begin{aligned}
f_{j,1}=\frac{\rho}{k_j}-\sum\limits_{l=1}^N\frac{\rho^2}{k_l}\frac{g_l}{\rho^2+k_jk_l}, \quad
f_{j,2}=1-\sum\limits_{l=1}^N\rho\frac{g_l}{\rho^2+k_jk_l},\\
h_{l,1}=1-\sum\limits_{j=1}^N\frac{\rho k_l}{k_j}\frac{g_j}{\rho^2+k_jk_l},\quad
h_{l,2}=\frac{k_l}{\rho}-\sum\limits_{j=1}^Nk_l\frac{g_j}{\rho^2+k_jk_l}.
\end{aligned}
\end{equation}
Thus,
\[\begin{aligned}
{\bf f}_1{\bf g}^T=\sum\limits_{j=1}^N\left(\frac{\rho}{k_j}-\sum\limits_{l=1}^N\frac{\rho^2}{k_l}\frac{g_l}{\rho^2+k_jk_l}\right)g_j,\\
{\bf h}_2{\bf \tilde{g}}^T=-\sum\limits_{l=1}^N\left(\frac{k_l}{\rho}-\sum\limits_{j=1}^Nk_l\frac{g_j}{\rho^2+k_jk_l}\right)\frac{\rho^2}{k_l^2}g_l,
\end{aligned}\]
which give rises to equation (\ref{d15}).

It is noted that equation (\ref{d11}) implies
another representation
\begin{equation}\label{d17}
v^2=-\frac{\det{\bf A}^{(b)}\det\tilde{\bf A}^{(b)}}{(\det{\bf A})^2}.
\end{equation}
Hence, we have a restraint condition about the discrete data
\begin{equation}\label{d18}
\rho^2=\frac{\det{\bf A}^{(b)}\det\tilde{\bf A}^{(b)}}{(\det{\bf A})^2}-i\partial_x\left(\frac{\det{\bf A}^{(a)}}{\det{\bf A}}\right).
\end{equation}

In particular, for $N=1$, we find
\[A=\tilde{A}=1-\frac{\rho^2g_1^2}{(k_1^2+\rho^2)^2}, \quad \det{{\bf A}^{(a)}}=-\frac{\rho}{k_1}g_1\left(1-\frac{\rho g_1}{k_1^2+\rho^2}\right).\]
\[\det{{\bf A}^{(b)}}=\left(1-\frac{\rho g_1}{k_1^2+\rho^2}\right)\left(\rho-\frac{k_1^2 g_1}{k_1^2+\rho^2}\right),\]
\[\det{\tilde{\bf A}^{(b)}}=\left(1-\frac{\rho g_1}{k_1^2+\rho^2}\right)\left(\rho-\frac{\rho^4}{k_1^4}\frac{k_1^2 g_1}{k_1^2+\rho^2}\right).\]

Then, the potentials can be given by
\begin{equation}\label{d18}
u=\frac{1}{4}\partial_x\ln\left(\frac{\rho-\frac{k_1^2 g_1}{k_1^2+\rho^2}}{\rho-\frac{\rho^4}{k_1^4}\frac{k_1^2 g_1}{k_1^2+\rho^2}}\right),
\end{equation}
\begin{equation}\label{b19}
v^2+\rho^2=i\partial_x\left(\frac{\frac{\rho}{k_1}g_1}{1+\frac{\rho g_1}{k_1^2+\rho^2}}\right).
\end{equation}
While, the solution (\ref{d17}) gives rise to the following representation
\begin{equation}\label{d20}
v^2=-\frac{\left(\rho-\frac{k_1^2 g_1}{k_1^2+\rho^2}\right)\left(\rho-\frac{\rho^4}{k_1^4}\frac{k_1^2 g_1}{k_1^2+\rho^2}\right)}
{\left(1+\frac{\rho g_1}{k_1^2+\rho^2}\right)^2}.
\end{equation}
We note that $g_{j,x}=-i(k_j+\rho^2/k_j)g_j$ by virtue of (\ref{c12}).
It is readily verified that the two forms of solution about $v^2$ are coincident with each other.

Let
\begin{equation}\label{d21}
\frac{k_1}{\rho}={\rm e}^{\epsilon_1}, \quad \frac{g_1}{\rho}={\rm e}^{-2i\theta_1},
\end{equation}
where
\[\theta_1=\rho\cosh\epsilon_1(x+\rho\sinh\epsilon_1\cdot t)+z_0.\]
Here $z_0$ is a constant. Then we have
\begin{equation}\label{d22}
u=\frac{1}{4}\partial_x\ln\left(\frac{2\cosh\epsilon_1-{\rm e}^{\epsilon_1}{\rm e}^{-2i\theta_1}}{2\cosh\epsilon_1-{\rm e}^{-3\epsilon_1}{\rm e}^{-2i\theta_1}}\right),
\end{equation}
\begin{equation}\label{d23}
v^2+\rho^2=i\rho\partial_x\left(\frac{2\cosh\epsilon_1{\rm e}^{-2i\theta_1}}{2{\rm e}^{\epsilon_1}\cosh\epsilon_1+{\rm e}^{-2i\theta_1}}\right).
\end{equation}
From (\ref{d21}), we know that the solitonic solution is invariant with respect to shifts in time $t$ and space $x$
by renormalizing the constant $z_0$.

In particular, we consider the case $k_1=i\kappa, (\kappa>\rho)$, and let
${\rm e}^{\epsilon_1}=i{\rm e}^\varepsilon, (\varepsilon>0)$, then the stationary soliton solution takes the form
\begin{equation}\label{d24}
u=\frac{1}{4}\partial_x\ln\left(\frac{\cosh(X)-\cosh(2\varepsilon)\cos(T)-i\sinh(2\varepsilon)\sin(T)}
{\cosh(X-2\varepsilon)-\cos T}\right),
\end{equation}
\begin{equation}\label{d25}
v^2+\rho^2=\frac{\rho{\rm e}^\tau}{2}\partial_x\left(\frac{{\rm e}^{iT}-{\rm e}^{X}}{\cosh(X)-\cos(T)}\right),
\end{equation}
where $\tau=\ln(2\sinh\varepsilon)$ and
\[X=2\rho\sinh\varepsilon\cdot x-\tau-\varepsilon+x_0,\quad T=\rho^2\sinh(2\varepsilon)\cdot t+t_0.\]
Here, we choose the solution is centered at $x=0$, or $z_0=0$. This solution is periodic in time, and its oscillation period is
\[{\cal{T}}=\frac{2\pi}{\rho^2\sinh(2\varepsilon)}.\]
Note that
\[\begin{aligned}
&{\cal{T}}\to\infty, \quad \varepsilon\to0+0, \quad |k_1|\to\rho,\\
&{\cal{T}}\to0, \qquad \varepsilon\to\infty, \quad |k_1|\to\infty.
\end{aligned}\]
Hence, the solution can be regarded as the Kuznetsov type solution \cite{spd22-507}.
One may find that, for any fixed value of $t$, the maximum of the solution occurs as
$X=\tau+\varepsilon$ for $v$ and $X=\tau+\varepsilon$ for $u$.
It is noted that the points at $(X-2\varepsilon)^2+(T+2m\pi)^2=0, (m=0,\pm1,\pm2,\cdots)$
are removable singularities of the $u$ and the points at $X^2+(T+2m\pi)^2=0, (m=0,\pm1,\pm2,\cdots)$ are
removable singularities of the solution $v^2$.
The typical behavior of the Kuznetsov-Ma type solution is given in figure 2.
 \begin{figure}[h]\
\includegraphics[width=6cm,height=6cm]{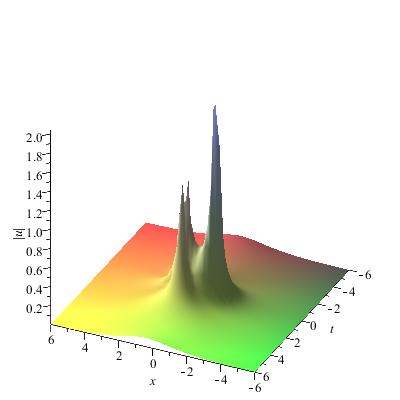}
\includegraphics[width=6cm,height=6cm]{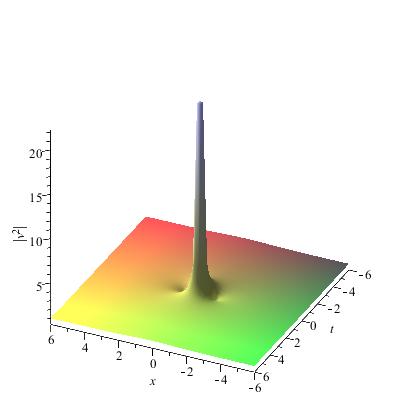}
\caption{\scriptsize Kuznetsov-Ma solution $|u|$ \eqref{d24} and $|v^2|$ \eqref{d25} with the parameters
chosen as $\rho=1, \varepsilon=\frac{1}{2}\ln2$. }
\end{figure}

Taking the limit $\varepsilon\to0$ or $k_1\to i\rho$ and choosing the proper origin of the soliton,
one obtains the rational solution
\begin{equation}\label{d25a}
\begin{aligned}
u(x,t)&=\frac{1}{4}\partial_x\left(\ln\frac{\rho^2x^2+\rho^4t^2-1-2i\rho^2t}{(\rho x-1)^2+\rho^4t^2}\right),\\
v^2(x,t)+\rho^2&=\partial_x\left(\frac{i\rho t- x}{ x^2+\rho^2t^2}\right).
\end{aligned}
\end{equation}
We note that the points at $(\rho x-1)^2+\rho^4t^2=0$ for $u$ and $ x^2+\rho^2t^2=0$ for $v^2$ are
removable singularities. The typical behavior of the rational solution is given in figure 3.
 \begin{figure}[h]\
\includegraphics[width=6cm,height=6cm]{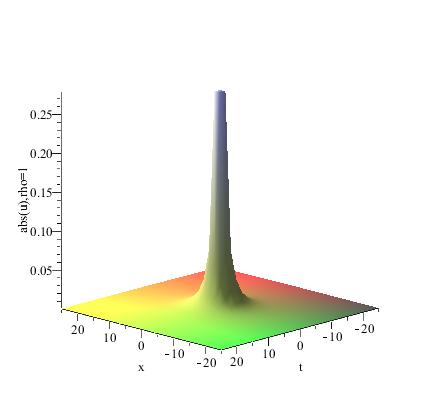}
\includegraphics[width=6cm,height=6cm]{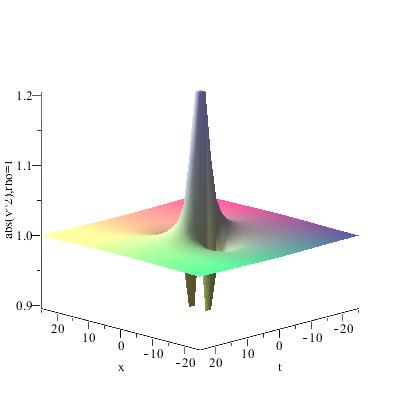}
\caption{\scriptsize Rational solution $|u|$ and $|v^2|$ in \eqref{d25a} with the parameters
chosen as $\rho=1$. }
\end{figure}

While, in the case $k_1=-i\tilde{\kappa}, (0<\tilde\kappa<\rho)$, and let ${\rm e}^{\epsilon_1}=-i{\rm e}^{\tilde\varepsilon}, (\tilde\varepsilon<0)$,
we have
\begin{equation}\label{d26}
u=\frac{1}{4}\partial_x\ln\left(\frac{\cosh(\tilde{X})
+\cosh(2\tilde\varepsilon)\cos(\tilde{T})-i\sinh(2\tilde\varepsilon)\sin(\tilde{T})}
{\cosh(\tilde{X}-2\tilde\varepsilon)+\cos \tilde{T}}\right),
\end{equation}
\begin{equation}\label{d27}
v^2+\rho^2=\frac{\rho{\rm e}^{\tilde\tau}}{2}\partial_x\left(\frac{-{\rm e}^{-i\tilde{T}}-{\rm e}^{\tilde{X}}}{\cosh(\tilde{X})+\cos \tilde{T}}\right),
\end{equation}
where $\tilde\tau=\ln(-2\sinh\tilde\varepsilon)$ and
\[\tilde{X}=-2\rho\sinh\tilde\varepsilon\cdot x-\tilde\tau-\tilde\varepsilon,\quad \tilde{T}=-\rho^2\sinh(2\tilde\varepsilon)\cdot t.\]
The properties of the solution can be discussed similarly. It is noted that, as $\tilde\varepsilon\to 0$ or
$k_1\to-i\rho$, the solution in (\ref{d26}) and (\ref{d27}) gives the same rational solution as (\ref{d25a}).

In the following, we consider another type of particular solution of the TD equation.
This solution is periodic in space and localized in time.
To this end, we consider the case that $k_1$ is located on the upper half-circle $|k|=\rho$ except the points $k=i\rho$,
and set $\epsilon_1$ in (\ref{d21}) as $\epsilon_1=i\omega$, then
\begin{equation}\label{d28}
\frac{g_1}{\rho}={\rm e}^{T_1+iX_1}, \quad T_1=\rho^2\sin(2\omega)t, \quad X_1=-2\rho\cos(\omega)x.
\end{equation}
Since $\cos\omega$ changes sign as $\omega\in[0,\pi]$, we express the solution in the following two forms
\begin{equation}\label{d29}
\begin{aligned}
u&=\frac{1}{4}\partial_x\ln\Omega_1, \quad v^2=-\rho^2+i\cos(\omega)\partial_x\Xi_1, \\
\Omega_1&=\frac{\cosh(T_1+\tau_1)\cos(2\omega)+i\sinh(T_1+\tau_1)\sin(2\omega)-\cos(X_1-\omega)}
{\cosh(T_1+\tau_1)-\cos(X_1-\omega)},\\
\Xi_1&=\frac{{\rm e}^{T_1+\tau_1}+{\rm e}^{i(X_1-\omega)}}{\cosh(T_1+\tau_1)+\cos(X_1-\omega)}, \quad (\omega\in[0,\frac{\pi}{2})),
\end{aligned}
\end{equation}
and
\begin{equation}\label{d30}
\begin{aligned}
u&=\frac{1}{4}\partial_x\ln\Omega_2, \quad v^2=-\rho^2-i\cos(\omega)\partial_x\Xi_2, \\
\Omega_2&=\frac{\cosh(T_1-\tau_2)\cos(2\omega)+i\sinh(T_1-\tau_2)\sin(2\omega)+\cos(X_1-\omega)}
{\cosh(T_1-\tau_2)+\cos(X_1-\omega)},\\
\Xi_2&=\frac{{\rm e}^{T_1-\tau_2}+{\rm e}^{i(X_1-\omega)}}{\cosh(T_1-\tau_2)+\cos(X_1-\omega)}, \quad(\omega\in(\frac{\pi}{2},\pi]),
\end{aligned}
\end{equation}
where $\tau_1=-\ln(2\cos\omega), (0\leq\omega<\pi/2)$ and $\tau_2=\ln(-2\cos\omega), (\pi/2<\omega\leq\pi)$.
In (\ref{d29}), $\sin(2\omega)>0, (0<\omega<\pi/2)$ (in $T_1$), we have the following asymptotic behaviors as $t\to\pm\infty$
\[\begin{aligned}
\Omega_1\to{\rm e}^{\pm2i\omega}, \quad \Xi_1\to H(t)=\{\begin{array}{cc}
1,&t>0,\\
0,&t<0.
\end{array}
\end{aligned}\]
While in (\ref{d30}), $\sin(2\omega)<0, (\pi/2<\omega<\pi)$,
\[\begin{aligned}
\Omega_2\to{\rm e}^{\mp2i\omega}, \quad \Xi_2\to H(-t), \quad t\to\pm\infty.
\end{aligned}\]
We note that $\Omega_1$ and $\Omega_2$ have the same asymptotic behaviors, so they can be regarded as homoclinic.
While the homoclinism for $\Xi_j$ is destroyed, but it can be modified to some extent after an action of differentiation.
In this sense, we also call the solution in (\ref{d29}) (or (\ref{d30})) the Akhmediev breather.
The typical behaviors of the Akhmediev breather are given in figure 4 and figure 5.
 \begin{figure}[h]\
\includegraphics[width=6cm,height=5cm]{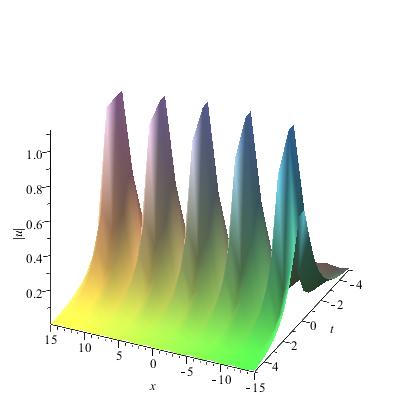}
\includegraphics[width=6cm,height=5cm]{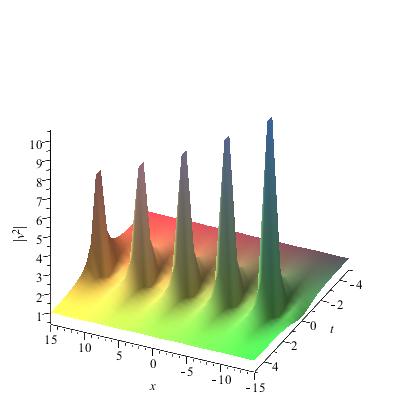}
\caption{\scriptsize Akhmediev breather from \eqref{d29}, obtained for $\rho=1, \omega=\frac{\pi}{3}$. }
\end{figure}
\begin{figure}[h]\
\includegraphics[width=6cm,height=6cm]{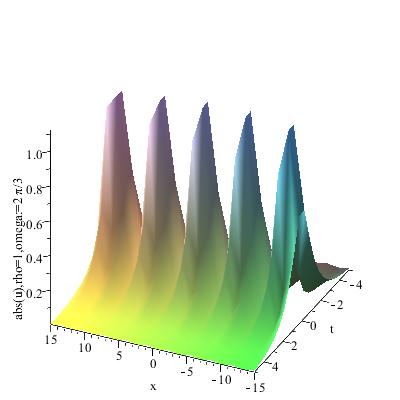}
\includegraphics[width=6cm,height=6cm]{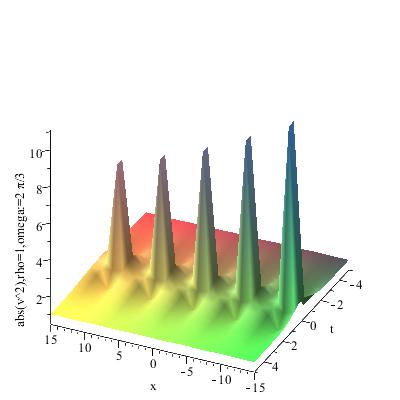}
\caption{\scriptsize Akhmediev breather from \eqref{d30}, obtained for $\rho=1, \omega=\frac{2\pi}{3}$. }
\end{figure}
We note that the dynamics of $u$ in (\ref{d29}) and (\ref{d30}) are same, but those of $v^2$ are different (see figure 6).
\begin{figure}[h]\
\includegraphics[width=6cm,height=5cm]{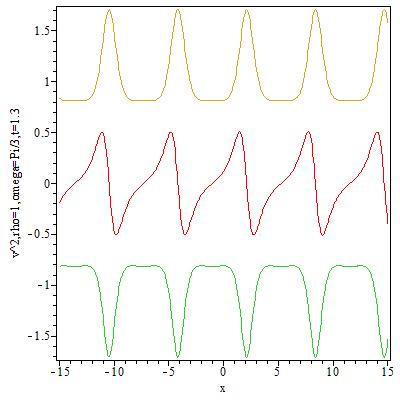}
\includegraphics[width=6cm,height=5cm]{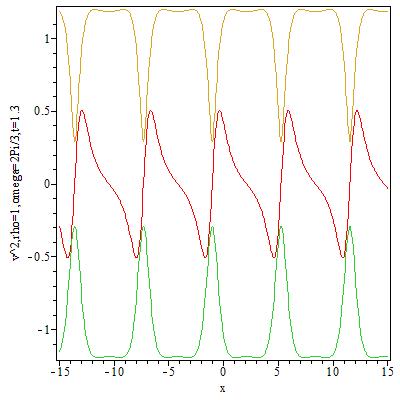}
\caption{\scriptsize Akhmediev breather for $v^2$, where $\rho=1, t=1.3$ and
left for $\omega=\frac{\pi}{3}$, right for $\omega=\frac{2\pi}{3}$.
Here, brown line--${\rm abs}(v^2)$, red line--${\rm Im}(v^2)$ and green line--${\rm Re}(v^2)$. }
\end{figure}

\section{Conclusion}
In this paper, we considered the TD equation with nonzero boundary condition by the inverse scattering transform method.
The Cauchy projectors were introduced to treat the inverse scattering problem. $N$-solitonic solution of the TD equation
was given by using the properties of the Cauchy type matrices. Note that the modulation instability is exist for
nonlinear integrable equation with nonzero boundary condition, and the TD equation has relation with
the Heisenberg spin equation, which is equivalent to the nonlinear Schr\"odinger equation.
We discussed the freak wave of the TD equation, and obtained the
Kuznetsov-Ma solution, rational solution and the Akhmediev breather.

\section*{Acknowledgments}
Project 11471295 was supported by the National Natural Science Foundation of China.


\end{document}